\documentclass[aps,prb,twocolumn,amssymb,floatfix,amsmath,aps,superscriptaddress,showpacs]{revtex4-1}
\usepackage{graphicx}
\usepackage{bm}
\usepackage[latin1]{inputenc}

\begin{document}

\title{Chaotic dynamics of magnetic domain walls in nanowires}

\author{A. Pivano}
\affiliation{Aix-Marseille Universit\'{e}, CNRS, IM2NP UMR7334, F-13397 Marseille
    Cedex 20, France}

\author{V.O. Dolocan}
\email{voicu.dolocan@im2np.fr}
\affiliation{Aix-Marseille Universit\'{e}, CNRS, IM2NP UMR7334, F-13397 Marseille
    Cedex 20, France}

\begin{abstract}
The nonlinear dynamics of a transverse domain wall (TDW) in Permalloy and Nickel nanostrips with two artificially patterned pinning centers is studied numerically up to rf frequencies. The phase diagram frequency - driving amplitude shows a rich variety of dynamical behaviors depending on the material parameters and the type and shape of pinning centers. We find that T-shaped traps (antinotches) create a classical double well Duffing potential that leads to a small chaotic region in the case of Nickel and a large one for Py. In contrast, the rectangular constrictions (notches) create an exponential potential that leads to larger chaotic regions interspersed with periodic windows for both Py and Ni. The influence of temperature manifests itself by enlarging the chaotic region and activating thermal jumps between the pinning sites while reducing the depinning field at low frequency in the notched strips. 

\end{abstract}
\pacs{75.60.Ch, 75.10.Hk, 75.40.Mg}

\date{\today}
\maketitle

\section{INTRODUCTION}

Nonlinear phenomena are studied in various fields ranging from atmospheric models to biophysics\cite{Moon,Hilborn}. In magnetism, the nonlinear dynamics has developed with the study of spin-wave instabilities in magnetic spheres and the nonlinear motion of a single spin due to the nonlinearity of the Landau-Lifshitz equation\cite{Wigen,Bertotti,Alvarez,Bragard,Smith}. The emergence of chaotic motion in a single Bloch domain wall (DW) was also studied under applied magnetic field in magnetic bubble garnet thick films\cite{Kosinski,Sukiennicki}. Recent results on non-linear DW dynamics discuss the current-driven motion of a confined DW in a constriction in relation to the DW spin-torque oscillators\cite{Matsushita}, the dynamic resonant response of magnetostatically interacting DWs in parallel nanowires\cite{OBrien} and the stochastic resonance of a DW between two pinning centers in a nanowire\cite{Martinez2011}.

The magnetic domains separated by DWs in nanowires (or nanostrips) are highly studied nowadays for storage applications like the racetrack memory\cite{Parkin} or for logic devices\cite{Allwood}. The operation of the racetrack memory is based on the synchronous displacement by applied field or current of a series of DWs that are normally pinned at constrictions or other patterned traps. Several types of traps and nanowires (cylindrical or strip) were studied\cite{Petit1,Petit2,Martinez2009,Dolocan1} along with the possible interaction between the DWs\cite{Pivano,Dolocan2}. The traps create a pinning potential for the DW that depends on the DW type (vortex or transverse). In low width/diameter nanowires the transverse DW represents the stable state that can be pinned in the potential well created by the traps. Until now, the rf driven DW dynamics needed to operate at high velocity the possible devices was not studied in these systems even it is well known that the motion between potential wells can lead to chaos. The presence of chaos in devices is normally not desirable as affecting performance and different methods can be employed in electrical systems to render the systems response periodic. From a fundamental point a view, the study of the chaotic movement of a magnetic DW contributes to the understanding of chaos universal features and of the nonlinear magnetization dynamics.

\begin{figure}[!t]
  \includegraphics[width=6cm]{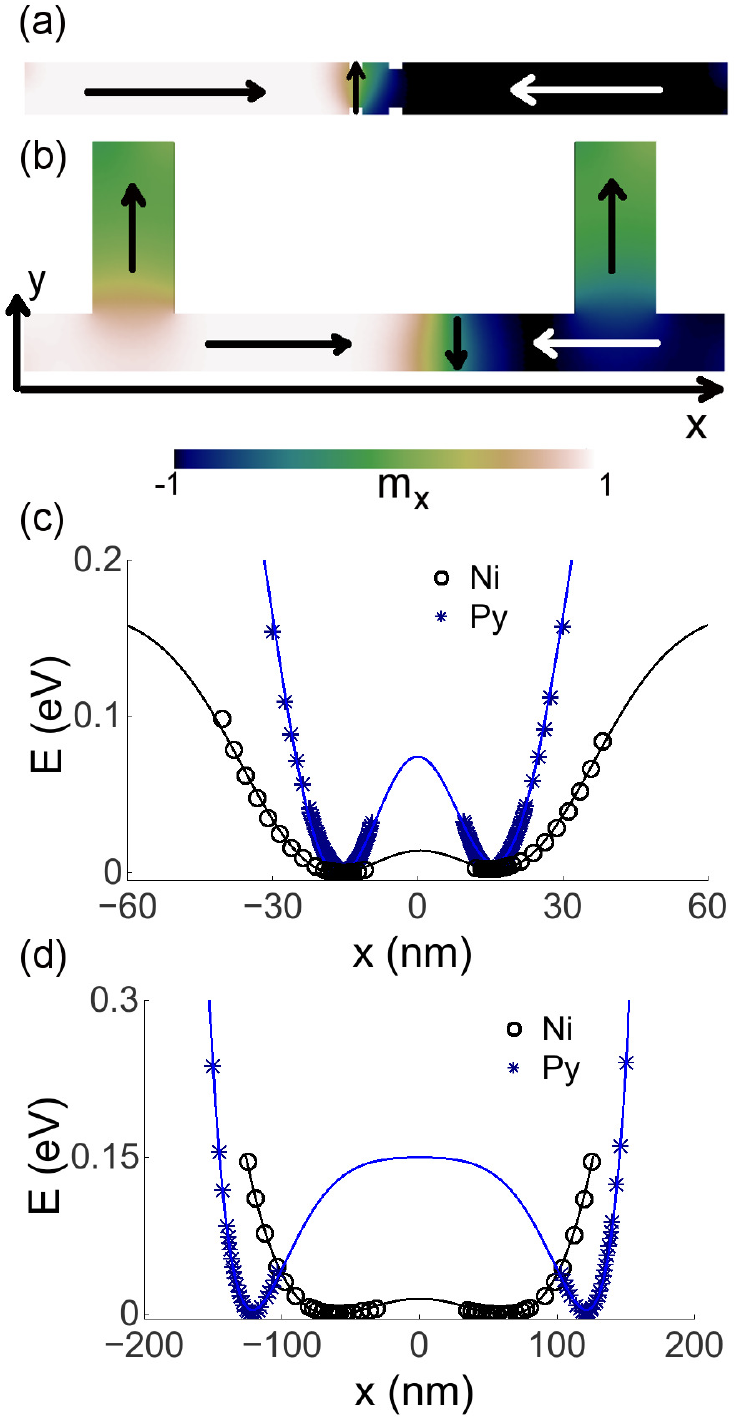}\\
 \caption{\label{Fig.1} (Color online) Simulated structures: planar nanowire with two symmetric double notches (a) and with two symmetric antinotches (b). The equilibrium position of a pinned DW is shown in each case. The arrows indicate the
direction of magnetization. Normalized potential pinning energy for notches (c) and antinotches (d) is shown for two different materials (Ni and Py) as determined by micromagnetic simulations.  }
\end{figure}

In this article, we study  numerically the nonlinear dynamics of a harmonic driven TDW between two pinning sites in a nanostrip up to rf frequencies. We investigate two pinning systems: one with two symmetric antinotches (T-traps) and one with two symmetric double notches (Fig.~\ref{Fig.1}(a) and (b)). The pinning sites create a double potential well in both system types which leads to a complex nonlinear dynamics of the DW function of the rf applied field or current. We find that the antinotched strip leads to an well known Duffing potential, while for the notched strip a phenomenological potential was inferred\cite{Martinez2009}. The double well potential depends on the material parameters (Ni or Py), which results in different phase diagrams and chaotic regions. The temperature influences the dynamics of the DW in the nanostrip increasing globally the chaotic window and reducing the depinning field at low frequency. 

This article is organized as follows. In Sec.~\ref{Model}, we present the micromagnetic and the stochastic 1D model used to calculate the rf wall dynamics. In Sec.~\ref{Results}, we compute and investigate the phase diagram of the DW dynamics in the different systems presented at T=0K and room temperature. Discussion and concluding remarks are presented in Sec.~\ref{Discussion}.


\section{Model}\label{Model}

We study numerically the dynamics of domain walls in nanostrips of Py or Ni with two types of pinning sites: notches and antinotches. The dimensions of notches and antinotches were chosen to have similar potential barriers between the pinning sites for the same material. The strip has a cross section of L$_y \times$L$_z$ = 50$\times$5nm$^2$ (with antinotches) or 60$\times$5nm$^2$ (with notches). The antinotches have dimensions of l$_x \times$ l$_y \times$ l$_z$ = 70 $\times$ 150 $\times$ 5 nm$^3$ separated by 350nm. Fig.~\ref{Fig.1}(b) shows the equilibrium position of a head-to-head TDW in the antinotched strip using the parameters of Nickel: saturation magnetization M$_s$=477kA/m ($\mu_0$M$_{s}$=0.6T), exchange stiffness A = 1.05 $\times$ 10$^{-11}$J/m, damping parameter $\alpha$=0.02. The dimensions of the notches are 20 $\times$ 10 $\times$ 5 nm$^3$ and are separated by 40nm for Ni nanostrip while for Py dimensions of 15 $\times$ 8 $\times$ 5 nm$^3$ were used with 30nm separation distance.  Fig.~\ref{Fig.1}(a) displays the equilibrium position of a head-to-head TDW in the notched strip using the parameters of Py: saturation magnetization M$_s$=860kA/m ($\mu_0$M$_{s}$=1.08T), exchange stiffness A = 1.3 $\times$ 10$^{-11}$J/m, damping parameter $\alpha$=0.01. An AC magnetic field or current were applied along the $x$-axis up to a frequency of 3GHz.

The DW dynamics was computed using 3D micromagnetic simulations with the Nmag package\cite{Fischbacher} and with the one-dimensional DW model\cite{Slonc,Thiaville}. For the micromagnetic computations, the strips were discretized into a mesh with a cell size of 3nm, inferior to the exchange length ($\sim$5nm). The average position of the DW center ($x$) is extracted for each applied field (in the axial $x$ direction) along with the azimuthal angle ($\psi$) of magnetization in the yz plane. No magnetocrystalline anisotropy is considered and the temperature is set to T = 0K. The effect of the temperature is introduced using the stochastic 1D model. The 1D model of the DW supposes that the DW is rigid and gives a quasi-quantitative understanding of the motion of TDWs. The Langevin equations of motion of the DW are\cite{Boulle,Lucassen}:

\begin{align}
\label{eq1}
(1+\alpha^2)\dot{X} =& -\frac{\alpha\gamma\Delta}{2\mu_0M_s S}\frac{\partial E}{\partial X} + \frac{\gamma\Delta}{2}H_k\sin 2\psi \nonumber\\
& + \frac{\gamma}{2\mu_0M_s S}\frac{\partial E}{\partial\psi} + \eta_{\psi} - \alpha\eta_X\\ 
(1+\alpha^2)\dot{\psi} =& -\frac{\gamma}{2\mu_0M_s S}\frac{\partial E}{\partial X} -\frac{\gamma\alpha}{2}H_k\sin 2\psi \nonumber\\
& - \frac{\alpha\gamma}{2\Delta\mu_0 M_s S}\frac{\partial E}{\partial\psi} + \eta_X + \alpha\eta_{\psi}
\end{align}

\noindent where $X$ and $\psi$ are the position and azimuthal angle (in the $yz$ plane) of the DW, $\Delta$ the DW width, S the section of the wire, $\gamma$ the gyromagnetic ratio, $M_s$ the saturation magnetization, $H_k$ the DW demagnetizing field, $\alpha$ the damping parameter, $\eta_X$ and $\eta_{\psi}$ represent stochastic Gaussian noise with zero mean value and correlations $\langle \eta_i(t) \eta_j(t') \rangle = (2\alpha k_B T)/(\mu_0M_s\Delta S \gamma_0)\delta_{ij}\delta(t-t')$. $E$ is the potential energy of the DW that includes the internal energy, the Zeeman energy, the effects of current, the interaction energy with other DWs and the pinning energy. The pinning field associated with the pinning potential is given by $H_{pin}(x)=-\frac{1}{2\mu_{0}M_{s}S}\frac{\partial E_{pin}}{\partial X}$ and the DW width variation is given by $\Delta(t)=\Delta[\Psi(t)]=\pi\sqrt{\frac{2A}{\mu_{0}{M_{S}}^{2}\sin^{2}{\Psi}+\mu_{0}M_{S}H_{k}}}$. The potential pinning energy E$_{pin}$ and the equilibrium position in each system are determined from quasistatic micromagnetic simulations and are shown in Fig.~\ref{Fig.1}(c) and (d) for the notched and the antinotched nanostrip. $H_k$ and $\Delta$ are also estimated from the micromagnetic computations. The pinning potentials determined by fitting the micromagnetic results are of bistable Duffing-type for antinotched strips E$_{pin}$ = a$x^2$ + b$x^4$ (for Py an extend Duffing potential with a $x^8$ term was added). For notched strips an exponential phenomenological potential was used as before\cite{Martinez2009}:

\begin{equation}
\label{eq3}
E_{pin}(x)=V_{0}-V_{1} \left[ \exp\Bigl({-\frac{(x+x_{L})^{2}}{L^{2}}}\Bigr)-\exp\Bigl({-\frac{(x+x_{R})^{2}}{L^{2}}}\Bigr) \right]
\end{equation}

\noindent with $x_L$ and $x_R$ the centers of each pinning site, $L$ the length and $V_1$ the effective depth of pinning sites. For both types of pinning centers (notches and antinotches), the energy barrier between the pinning wells is controlled by the dimensions and distance between the centers. 

To determine the apparition of chaos in the driven-DW dynamics as the control parameters are varied, we computed point-by-point phase diagrams for all systems with the 1D DW model for a large frequency range up to 3GHz. A similar micromagnetic computation will require an enormous execution time. The control parameters are the amplitude and frequency of the ac applied field (or current). The range of field amplitude ($\leq$50 Oe) is chosen to have only viscous motion (low DW velocity, no precession) and/or inferior to the depinning field (for notched stripes). The frequency range ($\leq$3GHz) is chosen to be on the same order of magnitude with access or reading/writing time in possible magnetic memories based on DW ($\sim$nanosecond). To compare the results of 1D model with the results of micromagnetic simulations, we computed the bifurcation diagrams and the Poincar\'{e} sections showing strange attractors by both methods for all systems at fixed frequencies.


\section{Results}\label{Results}

We start presenting our results with the dynamics of a DW under an applied magnetic field at zero temperature. Afterwards, we will investigate the dynamic behavior under applied current and the influence of the temperature. 

\subsection{Frequency -- magnetic field phase diagram at T=0K}

To identify and characterize chaos quantitatively we calculated the largest Lyapunov exponent which underlines the dynamics of chaotic behavior\cite{Hilborn,Bragard}. Another type of quantifier which punctuates the geometric nature of trajectories in phase space, the fractal dimension (more precisely the correlation dimension), was also calculated in some special cases. The method of Lyapunov exponent is based on the exponential instability of nearby chaotic trajectories in phase space to variations in the initial conditions. Noting the difference in the nearby trajectories $\delta x_0$ (for each component of the phase space) at the initial time $t_0$ and $\delta x_n$ the difference at a later time $t_n$ the Lyapunov exponent is defined by (in the long-term limit):

\begin{equation}
\label{eq4}
\lambda = \frac{1}{n} \ln \left ( \frac{\delta x_n}{\delta x_0} \right )
\end{equation}

A positive average Lyapunov coefficient implies a chaotic behavior of the system. To characterize the strange attractors we also use the correlation dimension which gives the probability of finding two points in the same cell (of given radius). A noninteger value of the dimension implies a strange attractor.

\begin{figure*}[!t]
  \includegraphics[width=17.5cm]{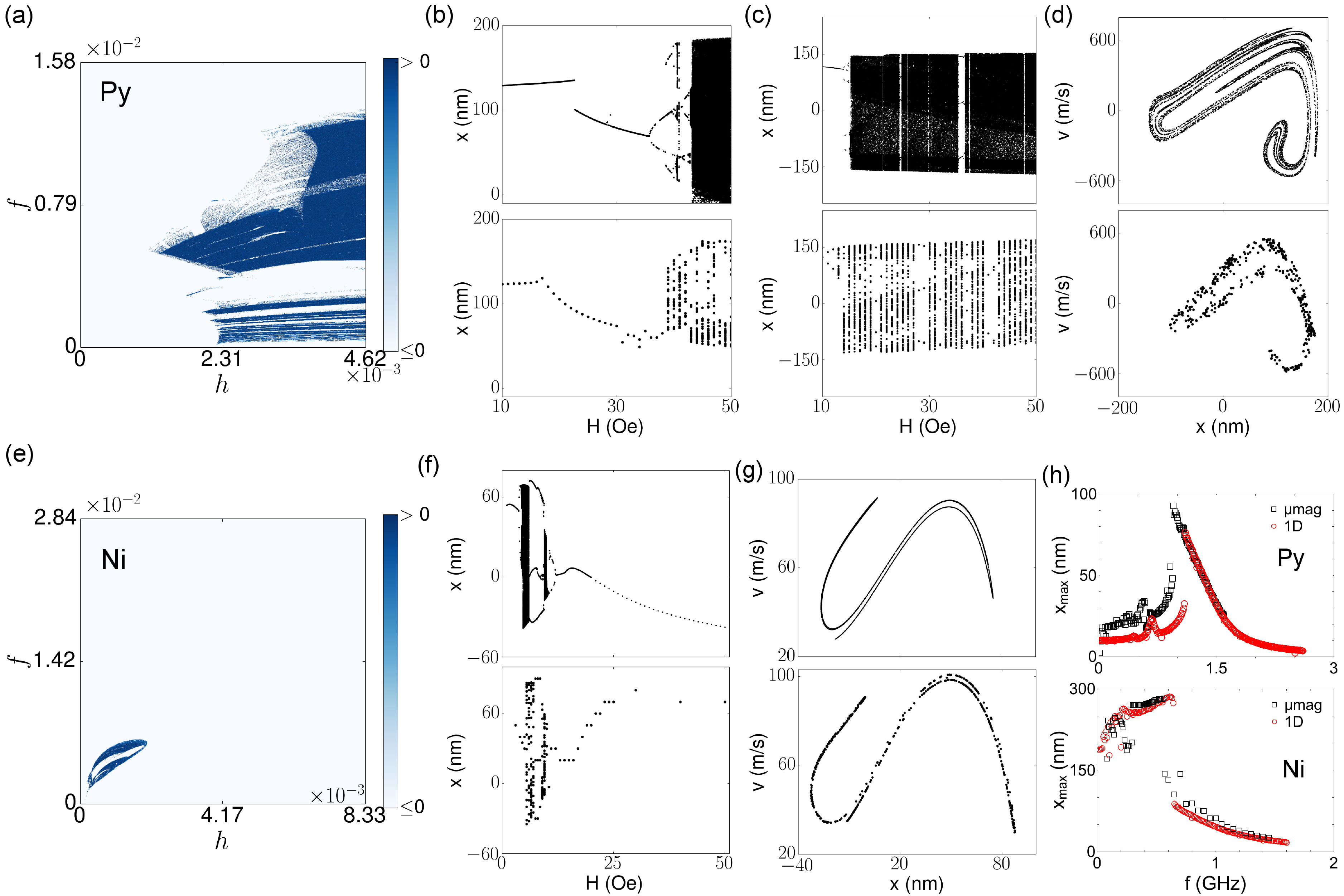}\\
 \caption{\label{Fig.2} (Color online) Normalized Lyapunov phase diagram for a DW at T=0K in a Py (a) and Ni (e) nanostrip with two symmetric antinotches. Dark color represents chaotic motion. The frequency $f$ is given in units of $\gamma_0$M$_s$, which corresponds to an absolute scale of 0-3GHz. The amplitude of the applied field $h$ is given in units of M$_s$, that corresponds to 0-50Oe in an absolute scale.  Bifurcation diagrams are shown in (b) and (c) computed at 500MHz and 1GHz respectively for Py. (d) Example of strange attractor obtained at H=50Oe from panel (c). (f) Bifurcation diagram for Ni at 500MHz. (g) Strange attractor for a field of H=6Oe from panel (f). The bifurcation diagrams and the strange attractor are computed with a 1D analytical model (upper panels) and by 3D micromagnetic simulations (lower panels). (h) Frequency response spectrum with jump phenomena for Py at H=10Oe and for Ni at H=20Oe.}
\end{figure*}

We start our study with a DW pinned in a nanostrip with two symmetric antinotches and submitted to a harmonic applied magnetic field. The asymptotic movement of the DW is equivalent to a classical Duffing oscillator. The phase diagrams of largest Lyapunov exponent in the parameter space frequency -- field amplitude are shown in Fig.~\ref{Fig.2}(a) and (e) for Py and Ni respectively at T=0K. The diagrams represent 1000$\times$1000 point-by-point integration with a fourth order Runge-Kutta scheme. A clear difference is observed between the two diagrams. In the Ni strip, we determine only a small window of chaos in the parameter space. The chaos appears at low fields ($\leq$ 12Oe) and frequencies inferior to 680MHz. Above these parameter values, the asymptotic motion of the DW remains periodic (period-1) between the two potential wells (below 1GHz) or in one potential well (above 1GHz). The periodicity of the asymptotic DW motion was confirmed micromagnetically for a number of cases at frequencies above 680MHz. A comparison between the micromagnetic and 1D model computed bifurcation diagram is shown in panel (f) for a frequency of 500MHz. The micromagnetic results display the same behavior as the 1D model. Chaotic windows appear between 4.5Oe and 10.2Oe with a large subharmonic period-3 motion in between. The transition to chaos starts with a period doubling bifurcation at 4.1Oe. The Poincar\'{e} section showing a strange (chaotic) attractor is shown in panel (g) for H=6Oe corresponding to the first chaotic window from (f) panel. The calculated correlation dimension (on 5000 points) of this attractor is 1.39.

In contrast, for Py antinotched strip, a large part of the parameter space is dominated by chaos. A periodic motion is determined for all frequencies at an applied field amplitude below 14Oe, and up to 50Oe for frequencies between 600MHz and 750MHz and above 2.5GHz. Up until 1.5GHz, the chaos starts at lower fields with increasing frequency. Two bifurcation diagrams are shown in panels (b) and (c) computed at 500MHz and 1GHz respectively. At 500MHz, the DW motion is periodic in one potential well until 22.5Oe, afterwards the motion stays periodic but takes place between the two potential wells. The onset of chaos starts with a period-doubling bifurcation at 35.7Oe and continues with a cascade of Feigenbaum-like period-doubling until 40.5Oe where it expands into a small window of chaos. At 41Oe, the 1D model predicts that the motion of the DW will revert to period-two motion with another cascade of period-doubling onset of chaos. Micromagnetically, the field-step used of 1Oe does not allow us to precisely identify the periodic window between 41 and 43 Oe. In general, the 1D model gives quantitatively the same results as the micromagnetic simulation (computed with less points) at low fields and quasi-quantitatively at high fields or high frequency. At 1GHz, the DW motion is dominated by chaos with many narrow windows where chaos reverts to periodic motion (period-3 or period-5). The Poincar\'{e} section for a strange attractor is shown in panel (d) for an amplitude of the applied field of 50Oe and a frequency of 1GHz. The correlation dimension of this attractor is 1.70.

Typical properties of anharmonic oscillations like jump and hysteresis phenomena were also determined (panel (h)) for the DW motion in the antinotched nanostrip. In the frequency response spectrum above a certain critical frequency, which depends only on the nonlinear terms in the potential and the magnetic damping, bistability occurs. For Ni nanostrip, the jump frequency varies between 0.31GHz at 6Oe to 0.62GHz at 20Oe. For Py, the jump occurs at 0.94GHz for H=10Oe and at 1.5GHz for H=30Oe. A secondary resonance is observed for Py at lower frequencies and is less evident for Ni.

\begin{figure*}[!t]
  \includegraphics[width=13cm]{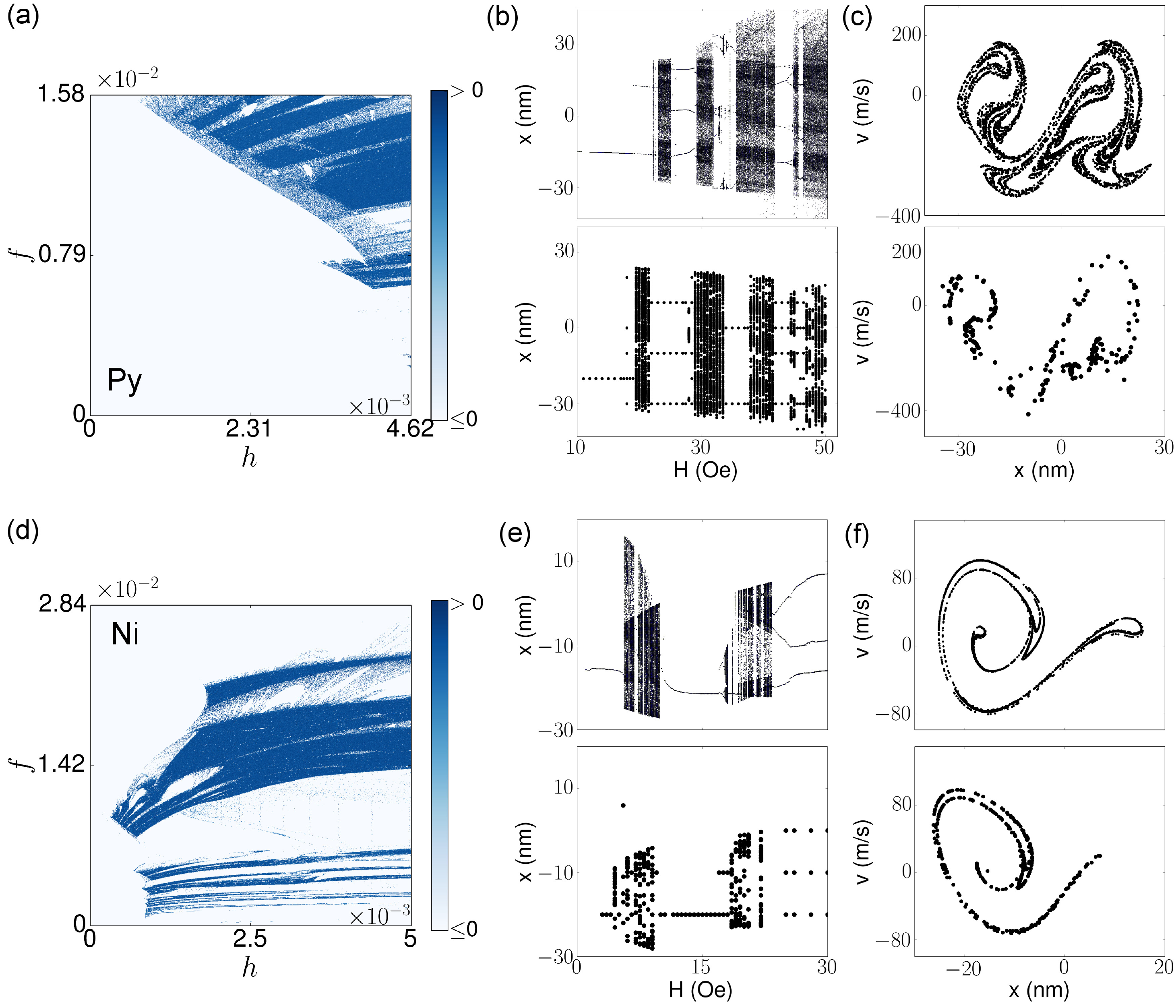}\\
 \caption{\label{Fig.3} (Color online) Normalized Lyapunov phase diagram for a DW at T=0K in a Py (a) and Ni (d) nanostrip with two double notches. Dark color represents chaotic motion. The frequency $f$ is given in units of $\gamma_0$M$_s$, which corresponds to an absolute scale of 0-3GHz. The amplitude of the applied field $h$ is given in units of M$_s$, that corresponds to 0-50Oe in an absolute scale for Py and to 0-30Oe for Ni. Bifurcation diagrams are shown in (b) and (e) computed at 2.5GHz for Py and 500MHz for Ni respectively. (c) Example of strange attractor obtained at H=30Oe from panel (b). (f) Strange attractor for a field of H=6Oe from panel (e). The bifurcation diagrams and the strange attractor are computed with a 1D analytical model (upper panels) and by 3D micromagnetic simulations (lower panels).}
\end{figure*}

The motion of the DW in the notched nanostrip is detailed in Fig.~\ref{Fig.3}. As two symmetric notches create a potential well that can pin a DW, the addition of another symmetric pair of notches at close distance to the first leads to a double well potential. The central barrier is highly dependent on this inter-notch distance and the potential has an exponential decay moving farther from the notch center. The external barriers are less important than for antinotches giving a depinning field around 50Oe (for antinotches H$_{depin}$=300Oe). For Py (panel (a)), the Lyapunov phase diagram shows a harmonic motion of the DW for all fields up to a frequency of 1.19GHz. This is in total contrast with the antinotch case where the periodic motion was obtained at high frequencies. Therefore, the control of the periodicity of the movement for a large range of frequencies can be obtained by changing the type of the pinning centers. For Py notched strip, chaos appears in the DW motion at high applied field amplitude and moves to lower field with increasing frequency. A bifurcation diagram computed at 2.5GHz is shown in panel (b). At low fields, the DW motion is periodic with only one stable attractor (in one potential well). At 19Oe the asymptotic DW motion is still period-1, but now there are two stable attractors each corresponding to the two stable points (potential wells). The DW motion remains periodic in one potential well. The onset of chaos is sudden with an abrupt transition from periodic motion. One can observe several chaotic windows interspersed with periodic windows. The periodic windows are larger than in the case of antinotches with the largest being the first window (period-3) of 4.1Oe (data from 1D model). The DW asymptotic motion observed just before the period-3 window shows intermittency (intervals of period-3 and chaotic motion). After the appearance of the period-3 window, the chaotic intervals become larger and more frequent. The third chaotic interval contains actually tiny periodic windows (mostly period-5). The Poincar\'{e} section for a strange attractor is shown in panel (c) at H=30Oe. The correlation dimension of this attractor is 1.84. 

For Ni notched nanostrip (panel (d)), the DW motion is periodic at all fields at low ($\leq$100MHz) or high frequencies ($\geq$2.75GHz) and between 500MHz and 750MHz. The chaotic regions contain large pockets of periodic regions. A detailed bifurcation diagram (panel (e)) computed at 500MHz shows large periodic windows together with chaotic intervals. The chaotic windows contain several tiny periodic windows. The onset of chaos appears like an abrupt transition from a period-1 motion inside one potential well. The second periodic interval is also of period-1 with the motion taking place between the two potential wells. At higher applied field, the motion is period-3.  The Poincar\'{e} section for a strange attractor is shown in panel (f) at H=6Oe. The correlation dimension of this attractor is 1.38.


\subsection{Frequency -- current phase diagram at T=0K}

\begin{figure}[!th]
  \includegraphics[width=5.5cm]{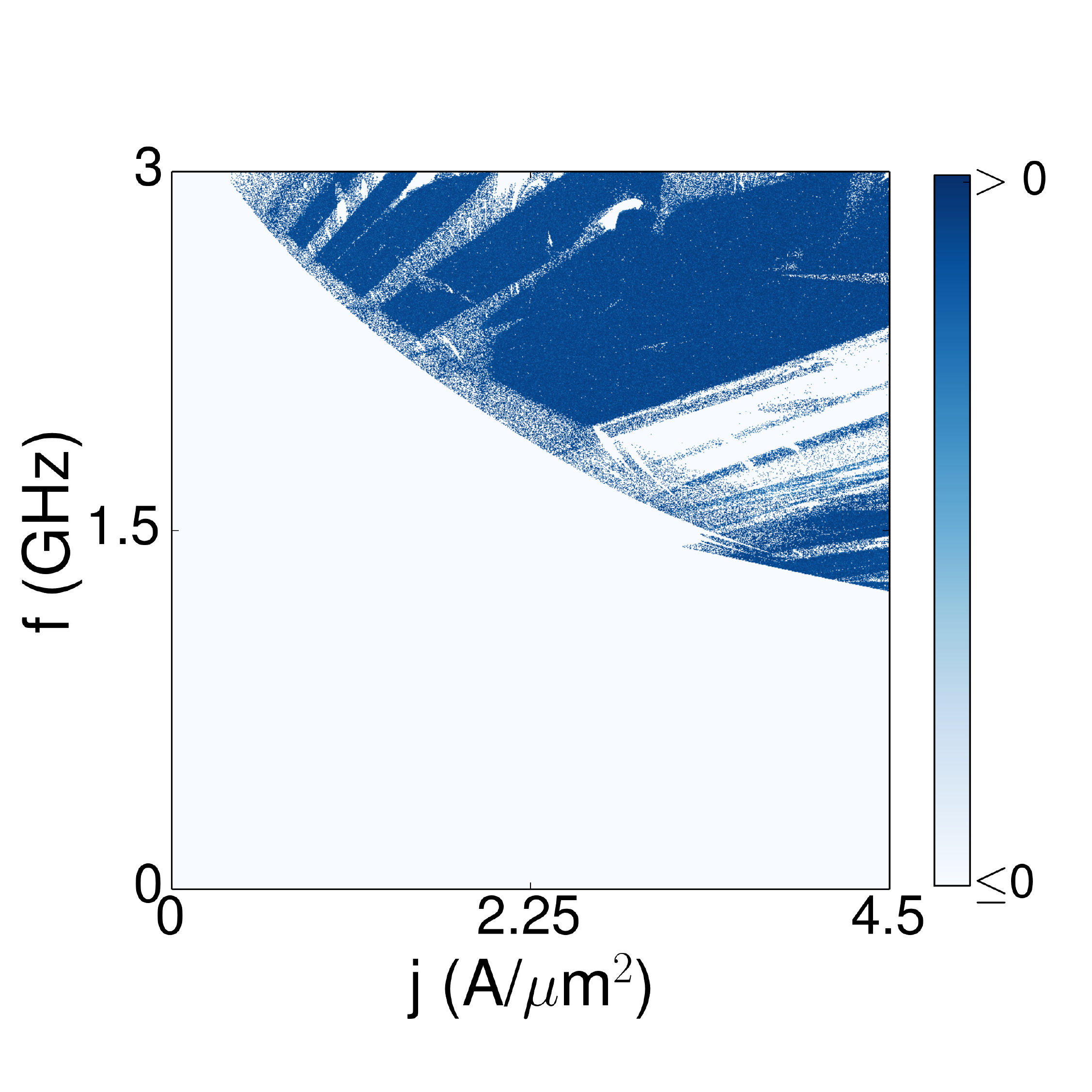}\\
 \caption{\label{Fig.4} (Color online). Frequency--current phase diagram for a DW at T=0K in a Py nanostrip with two double notches.}
\end{figure}

The effect of a spin-polarized rf-current (zero applied magnetic field) on the DW dynamics is shown in in Fig.~\ref{Fig.4} for the Py notched system at T=0K. The nonadiabatic spin-torque parameter $\beta$ that allows the DW to be driven in the dissipative regime is taken here equal to 2$\alpha$. There is still much debate over the actual value of $\beta$, with a consensus that is of the same order as the damping parameter $\alpha$. The regime chosen here, $\beta > \alpha$, corresponds to a similar behavior as the field-driven case for the velocity of the DW\cite{Lucassen} showing a Walker-breakdown behavior. Therefore, we expect a similar phase diagram as in Fig.~\ref{Fig.3}(a). In the two cases, field-driven and current-driven, we observed the same behavior, with a periodic motion until almost 1.5GHz, with chaotic motion appearing at higher frequencies. Same type of periodic windows inside the chaotic region are computed, the only difference being the slope of the boundary between the periodic and chaotic motion which depend on the value of the non-adiabatic spin-torque parameter.   

The frequency--current phase diagrams for the antinotched strip (Py and Ni) are similar with the ones presented in  Fig.~\ref{Fig.2} and are not shown.

\subsection{Temperature dependence}

The temperature influences the motion of the DW in two ways: on one hand the thermal noise disturbs the DW oscillation by modifying its shape and on the other hand the temperature activates jumps over the middle potential barrier at smaller applied field/current. These thermal activated jumps usually follow an Arrhenius-N\'{e}el type law\cite{OBrien2,Himeno,Lucassen} with the transition rate given by:

\begin{equation}
\Gamma = \Gamma_0 e^{-\Delta V/k_B T}
\end{equation}

\noindent where $\Delta V$ is the barrier height, and $\Gamma_0$ the attempt frequency at zero temperature. $\Gamma_0$ is estimated in the range 10$^7$-10$^{12}$Hz\cite{OBrien2,Himeno} and using an experimental time of 1ms\cite{OBrien2}, the potential barrier a DW could surmount varies from 0.23 to 0.51eV at room temperature. This give a high estimate of the transition rate, but is more than enough to activate jumps between the two potential wells presented in section~\ref{Model}.

Using Eq.~\eqref{eq1}, we determined the stochastic motion of the DW by computing 100 realizations of the DW motion for a finite number of points (15$\times$500) in the phase diagrams of Fig.~\ref{Fig.2}-\ref{Fig.3}. Globally, for antinotched and notched nanowires, the motion of DW at low fields (below 10-15Oe) is highly affected by thermal noise and the periodic motion can no longer be extracted. The motion is chaotic as the FFT of the DW motion shows noise for a large range of frequency. At higher fields, this thermal noise has lesser influence on the harmonic motion of the DW and periodic motion can be detected at 300K for some frequency range.

\begin{figure*}[!t]
  \includegraphics[width=17.5cm]{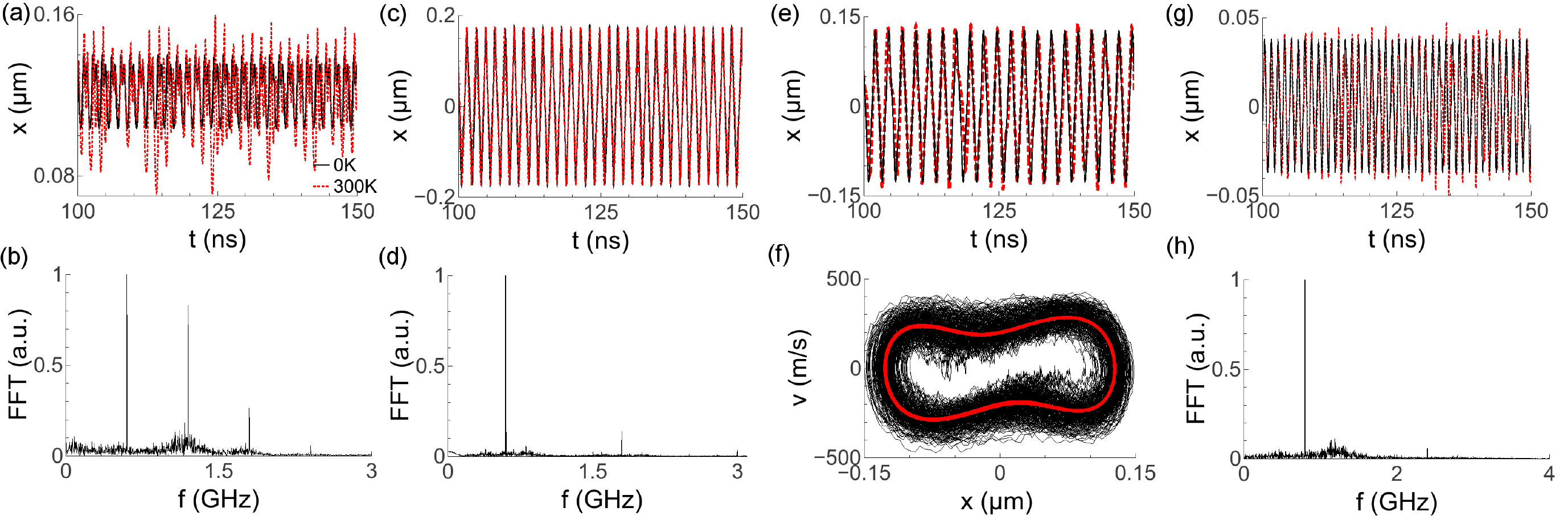}\\
 \caption{\label{Fig.5} (Color online). Comparaison between the DW motion at 0K (full line) and 300K (dotted line) for the Py antinotched strip (a) and (c), Ni antinotched strip (e) and Ni notched strip (g). The control parameters are: (a) 15.7Oe, 600MHz, (c) 22.6Oe, 600MHz, (e) 18Oe, 400MHz, (g) 17.2Oe, 800MHz. The power spectrum at 300K is shown in (b), (d) and (h) corresponding to (a), (c) and (g) respectively. (f) Phase-space plot of the DW motion from panel (e). The thick line corresponds to 0K.  }
\end{figure*}

Typical temperature modification of the harmonic motion of the DW are shown in Fig.~\ref{Fig.5}. The usual behavior at low fields is shown in panel (a) for the Py antinotched wire at 600MHz (H=15.7Oe). The DW motion takes place in one well (period-2) at 0K (continous line) with the FFT showing harmonics from 1 to 6 (image not shown). At 300K, the motion is still periodic in one well (period-2) with a jump probability to the second well of only 2$\%$. We observe in this case an amplitude modulation of the DW motion (dotted line) which reflects in the FFT that shows noise in a large range of frequency (panel (b)). The noise peak around 1.2GHz (which here superposes with harmonic 2) appears in all the simulations for the Py antinotched wire at 300K. To quantify the noise in the system, we calculated the area under the large noise peak (weight intensity of the peak) for different magnetic fields and frequencies. We found that the weight of the noise peak decreases exponentially with increasing magnetic field and increasing frequency up to 1.5GHz. This corresponds to the noise being larger when the oscillations take place in one well and the middle potential barrier is higher. As the amplitude of the applied field is increased the height of the barrier diminishes and the oscillations take place between the two wells. This is the case in panel (c), where the DW motion is shown for an applied field of 22.6Oe (600MHz, Py antinotched wire). The motion is period-1 between two wells (with 100$\%$ probability) with a very small amplitude modulation at 300K. The FFT of the signal shows the odd harmonics with very small noise around them (panel (d)). Therefore, the phase diagram of the Py antinotched wire is modified as follows: for all frequecies at low applied fields (inferior of 15Oe) the DW motion becomes chaotic due to thermal noise with the motion occuring in one well. Above 15Oe and up to 1.5GHz, the phase diagram does not change much with the periodic motion showing small thermal noise and chaos appearing at smaller fields. Above 1.5GHz, the DW motion occurs in one potential well for applied fields up to 50Oe. The noise is important and the signal is chaotic. The temperature activates jumps between wells at lower fields (for example at 20Oe for 2.2GHz), with the motion still occuring in one well. At higher fields, the jumps appear more often with intermittent oscillations between the two wells.

The phase diagram for the Ni antinotched wire (Fig.~\ref{Fig.2}(e)) changes more dramatically under the influence of temperature. Above 900MHz, for all applied fields the DW motion becomes chaotic (at 0K being periodic). Below 900MHz the DW motion is periodic for a large range of applied field, starting from 3Oe at 50MHz and from 7Oe at 200MHz. The periodic window diminishes with increasing frequency starting from 11Oe at 400MHz, 19Oe at 600MHz and 39.3Oe at 800MHz. An example of the temperature influence on the DW motion is shown in Fig.~\ref{Fig.5}(e)-(f) for the Ni antinotched wire. For these control parameters (H=18Oe, f=400MHz), the motion is periodic between the two wells at 0K, showing a periodic attractor in the phase-space (thick red line in panel (f)). To this periodic attractor it will correspond exactly one point in the Poincar\'{e} section in phase space. At 300K, the motion stays periodic with small noise (amplitude modulation) characterized by a periodic attractor which becomes enlarged consisting of several loops. Still the motion is periodic (or quasi-periodic) with the Poincar\'{e} section showing a cloud of points around the point corresponding to 0K. The FFT (not shown) is similar with the one in panel (d), showing harmonics 1, 3 and 5 with a small noise peak around 675MHz.

In the case of the notched wire, as the external potential barrier for depinning  (escaping from the potential) is six times smaller than the antinotched wire, the depinning field is strongly affected by the temperature\cite{Pivano}. For Py, the thermal noise at low fields (inferior of 15Oe) is large as in the case of the antinotched wire and the motion is chaotic. Between 15 and 25Oe at low frequency (below 800MHz), the DW motion is periodic in one well with moderate noise and nonzero probability of jump to the second well. The jump probability increases with the applied field and is 29$\%$ at 15Oe and 93$\%$ at 25Oe (500MHz). Above 25Oe and up to 1.5GHz the DW depins. The depinning probabilities at 800MHz are 1$\%$ at 15Oe, 39$\%$ at 20Oe and 88$\%$ at 25Oe. Above 800MHz, the noise is too important and the motion is chotic. Therefore, the only periodic window is below 800MHz at fields between 15 and 25Oe with the window slowly reducing with increased frequency.

For the notched Ni strip, the periodic window is larger than for Py at 300K. As the middle potential barrier is lower, the DW motion takes place between the two wells even at 10Oe (at 100MHz). The DW motion is periodic between 10 and 30Oe at 100-200MHz and between 15 and 25Oe up to 1.2GHz. Above 1.2GHz, the motion is chaotic between two wells with some small periodic windows (intermittency). A typical harmonic DW motion is shown in panel (g) of Fig.~\ref{Fig.5} for a field of 17.2Oe at 800MHz with small thermal noise at 300K (FFT in panel (h)).

\section{Discussion and conclusion}\label{Discussion}

The emergence of chaos in the DW motion depends largely on the damping parameter $\alpha$ and on the precise form of the potential and material parameters. For Ni, a value of $\alpha$=0.02 was used which corresponds to the calculated value at room temperature. At low temperature, below 80K, the damping parameter increases for Ni up to a value of 0.1\cite{Gilmore}. If an actual value of $\alpha$ = 0.1 is used, the phase diagram changes and for the antinotched strip only periodic motion of the DW is determined (no small chaotic region). For the notched strip, the chaotic region diminishes to a small region (inclined bubble-like) that stretches between 800MHz and 1.5GHz with the largest width of 10Oe (figure shown in Ref.\onlinecite{SM}). This does not influence the behavior at room temperature, where the periodic motion is observed only up to 900MHz (antinotched wire) and up to 1.2GHz (notched wire). Similarly, for Py strip with notches, if the value of $\alpha$ is increased from 0.01 to a value of 0.1 the DW motion is almost completely periodic with a very small chaotic region around 2GHz and above 45Oe.

The influence of the saturation magnetization on the DW motion is more complex as the pinning energy (and demagnetization energy) and the DW width vary. For the nanostrip with notches, varying M$_s$ will change slightly the curvature of the potential and the central potential barrier height, which increases with M$_s$\cite{SM}. When M$_s$ is decreased the DW movement will be more chaotic at high fields (more jumps between wells) and chaos will appear at slightly lower frequencies (above 40Oe). At higher M$_s$, the chaotic motion moves to higher frequencies and larger periodic windows are determined inside the chaotic area. An example of this effect for Py notched strip is shown in Ref.\onlinecite{SM}. However the dependence is not linear, and the changes in the phase diagram are small. In the case of the antinotched strip, the variation of the M$_s$ will change more drastically the form of the potential well (and equilibrium positions) and for values too far apart from the actual value (going to 1000kA/m from 860kA/m for Py) the potential changes too much and cannot be fitted anymore to a classic Duffing potential. Globally, the same behavior is expected with apparition of chaos at lower frequencies and fields.  

Varying the exchange constant A will modify mostly the potential barrier height for the notched strip, while the curvature of the potential stays almost the same. The chaotic region is displaced to higher frequencies and fields when A is diminished, with changes inside the chaotic region as the pockets of periodic motion increase. Therefore, in principle, diminishing the exchange constant and increasing the saturation magnetization should displace the chaotic region to higher fields and higher frequencies.

In conclusion, we have shown that the dynamics of a TDW between two artificial pinning centers in a nanostrip can have a complex behavior. We have found that for certain parameter range, the DW motion is dominated by chaos or shows intermittency. At low temperature, there is a large region of periodic motion of the DW under applied field or current up to 3GHz and 50Oe. For Py, changing the type of pinning centers from antinotches to notches results in a high frequency to low frequency periodic DW motion, while for Ni the antinotches result in almost periodic DW motion. As typical experimental observations are realized at room temperature (for example by electrical detection of the DW motion), our stochastic results predict that periodic DW motion occurs only below 1.2-1.5GHz for all types of pinning centers (for Py and Ni). The smaller region of periodic DW motion is predicted for Py notched nanostrip, while the largest for Ni antinotched strip. In general, the Ni notched and antinotched wires seem better suited to be used at room temperature up to GHz frequencies for periodic displacement of DWs.

The chaotic dynamics is not less important in these systems with the possibility of synchronization of chaotic DW signals that could have potential applications in secure communications\cite{Peterman}. A future direction to be explored is chaos control in these systems which could have potential applications\cite{Chacon,Meucci}. Our analysis could apply to other similar systems where chaos may appear as vortices in superconductors\cite{Olive} or skyrmions in magnetic nanostructures\cite{Sampaio}.


\begin{acknowledgments}
This work was granted access to the HPC resources of Aix-Marseille Universit\'{e} financed by the project Equip$@$Meso (ANR-10-EQPX-29-01) of the program "Investissements d'Avenir" supervised by the Agence Nationale pour la Recherche. 
\end{acknowledgments}


\end{document}